\documentclass[%
pra,
twocolumn,
reprint,
amsmath,
amssymb,
aps]{revtex4}

\usepackage{natbib}
\usepackage{graphicx}
\usepackage[tight]{subfigure}
\usepackage{bm}
\usepackage{dcolumn}
\usepackage{amssymb}
\usepackage{amsbsy}
\usepackage{times}
\usepackage{color}
\usepackage{float}
\usepackage[colorlinks,bookmarks=false,citecolor=blue,linkcolor=red,urlcolor=blue]{hyperref}

\begin{document}

\title{Phase diagram of disordered Bose-Hubbard model based on mean-field and percolation analysis in two dimensions and at fixed $n=1$ filling}

\author{Manjari Gupta}

\affiliation{Harish-Chandra Research Institute, HBNI, Chhatnag-Road, Jhunsi, Allahabad-211019, India}

\date{\today}

\begin{abstract}

We present a phase diagram of Bose-Hubbard model with on-site chemical potential disorder at two dimensions within the scope of mean-field theory. The phase diagram in the disorder strength ($\Delta$) and the on-site repulsion ($U$) for disordered BHM at fixed filling $\langle{n\rangle}=1$, show interesting re-entrance of superfluid phase, sandwiched between Bose-glass phases, as observed by the previous QMC results. We probe the Bose-glass to superfluid transition, as a percolation transition, based on the mean-field results at various parts of the phase diagram using both $\Delta$ and $U$ as the tuning parameter. We argue the robustness of the re-entrant superfluid.
\end{abstract}

\maketitle

\section{Introduction}

Bose Hubbard model with chemical potential disorder is a useful model to study disorder physics in presence of strong interaction. Although there are a number of ways to introduce disorder, this is simple as well as easy to implement in ultra-cold experiments \cite{DeMarco,Bruce2016}. This model was first introduced in the famous paper by Fisher {\it et. al.} (Ref. [\onlinecite{Fisher}]) exploring the fate of the Mott and superfluid phases in the presence of disorder. With the introduction of diagonal disorder, along with the superfluid and Mott phase there is a Bose-glass phase that appears (Ref. \cite{Fisher,3DdbhmQMC, 2DdbhmQMC}). Both superfluid and bose-glass are gapless compressible phases with non-zero superfluid stiffness in the superfluid phase and zero in the bose glass phase. The simplest form of the disorder distribution is the box disorder where the disordered part of the chemical potential ($\epsilon_i$) has a uniform distribution within a bound of width $\Delta$ ($[-\Delta/2,\Delta/2]$). For a bound type of disorder (as the box disorder) the `theorem of inclusion' compels the MI to SF transition to be through a Bose-glass phase \cite{3DdbhmQMC,QuasiOneD,Troyer2009}. Hence, the superfluid to Mott transition is completely replaced by superfluid to BG transition and BG to Mott transition.  

The phase diagram of Disordered Bose-Hubbard model (DBHM) at any fixed integral filling, (typically $n=1$ filling) are of recent interest. There are QMC studies \cite{3DdbhmQMC, 2DdbhmQMC,Ceperley,Nandini1993,MG_qmc,scalling_qmc_1,Ranjan} as well as analytical studies \cite{mfT_not1,Smft_1,Smft_2} aiming at $\langle{n\rangle}=1$ phase diagram for DBHM as a function of interaction strength ($U$) and disorder strength ($\Delta$) at both two and three dimensions. The re-entrant superfluid phase is of interest. This happens for the the cases where either of $U$ ($>U_c$) or $\Delta$ is kept fixed and the other one is varied. There are extensive QMC studies \cite{3DdbhmQMC, 2DdbhmQMC}  using worm algorithms aim at the ground state phase diagram at fixed filling. Among analytical approaches, stochastic mean-field and Gutzwiller mean-field studies  \cite{mfT_not1,Smft_1,Smft_2,PRA031601} aiming both the ground state and finite temperature phase diagram of the aforementioned phase diagram for fixed $n=1$ filling. Although the simulational studies so far investigated the fixed filling phase diagram to some extent, more analytical studies are needed to develope better understanding of the physics. Analytical techniques based on a strong-coupling mean-field approach can give a good starting point for that perpose.

In this paper we present mean-field calculations for the disordered BHM and a straightforward technique to distinguish all three phases. We also demonstrate how mean-field calculations can serve as a starting point for exploring the phenomenology regarding the phase diagram in question. The paper is organised as follows, we first describe the disordered BHM and the strong-coupling mean-field technique used, then we discuss the robustness of the re-entrant superfluid phase (at strong-coupling), which is of current interest, within mean-field analysis. Followed by that we discuss determination of the BG to superfluid transition in terms of classical percolation and the results obtained from that.

\section{Bose-Hubbard Hamiltonian with diagonal disorder}

The Bose-Hubbard model with an onsite-repulsive term and nearest neighbour hopping for the bosons has been a useful base to study strongly correlated many-body physics. Adding a simple disorder term in the chemical potential gives rise to a rich phase diagram as observed in the quantum Monte-carlo calculations.

\begin{equation}
\mathcal{H}=-t\sum_{\langle{ij\rangle}}(b^{\dagger}_ib_j+h.c.)+\frac{U}{2}\sum_i n_i (n_i-1)-\sum_i (\mu+\epsilon_i)n_i
\label{eq1}
\end{equation}

$b_i$ ($b_i^{\dagger}$) is the annihilation (creation) operator for the bosons, $n_i$ is the number operator at site $i$. $t$ is the kinetic energy expense for boson hopping to the nearest neighbour sites, $U$ is the on-site repulsive interaction and $\mu$ is the chemical potential. $\epsilon_i$'s are chemical potential disorder at each site derived from a box distribution defined in $-\Delta/2 \le \epsilon_i \le \Delta/2$, where $\Delta$ is the disorder strength.

\subsection{Strong-coupling Mean-field of DBHM}

We perform a simple mean-field calculation at zero temperature to explore the $\Delta-U$ phase diagram. We use a mean-field decoupling of the hopping term in Eq. \ref{eq1} and calculate $\phi_i$ self-consistently (at each site $i$) by diagonalizing the single site hamiltonian ($h_i$) at each site as given below.

\begin{equation}
\mathcal{H}=\mathcal{H}_0+\mathcal{H}_1
\end{equation}

\begin{equation}
\mathcal{H}_0=\sum_{i}h_i+t^{-1}\sum_{\langle{ij\rangle}}(\phi_i^*\phi_j+\text{c.c.}) 
\end{equation}

\begin{equation}
h_i=-t(b^{\dagger}_i\phi_i+b_i\phi_i^*)+\frac{U}{2}n_i (n_i-1)-(\mu+\epsilon_i)n_i
\label{eq2}
\end{equation}

Where,

\begin{equation}
\phi_i=-t\sum_{j \text{nn to }i}\langle{b_j\rangle}.
\label{eq6}
\end{equation}

We drop the fluctuation term ($\mathcal{H}_1=-t\sum_{\langle{ij\rangle}}[(b^{\dagger}_i-\langle{b_i\rangle}^*)(b_j-\langle{b_j\rangle})+\text{h.c.}]$). The expectation value of the annihilation operator ($\langle{b_i\rangle}$) is carried out self-consistently at each site $i$, w.r.t. $h_i$. 

Without the disorder the superfluid phase is easily distinguishable from the Mott phase as the expectation value of the annihilation (creation) operator $\langle{b\rangle}$ ($\langle{b^{\dagger}\rangle}$) is non-zero in the superfluid phase.  In the presence of disorder, a Bose-glass phase emerges with non-zero $\langle{b\rangle}$. It is well accepted in the literature that the Bose-glass region usually consists of superfluid puddles embedded in Mott background (Ref. \cite{3DdbhmQMC,Ceperley,Droplet2008}). Hence, in the presence of finite disorder ($\Delta$) and interaction strength ($U$), the disorder averaged order parameter is never exactly zero in the Bose-glass phase for a thermodynamically large system. Existing methods of distinguishing BG and superfluid phase include compressibility, superfluid stiffness and Edward Anderson's order parameter \cite{EdwardAndersons}. All the three are beyond the scope of our mean-field approach as discussed above as these are essentially response functions non-local in time. 

\begin{figure}[t!]
\includegraphics[width=9cm]{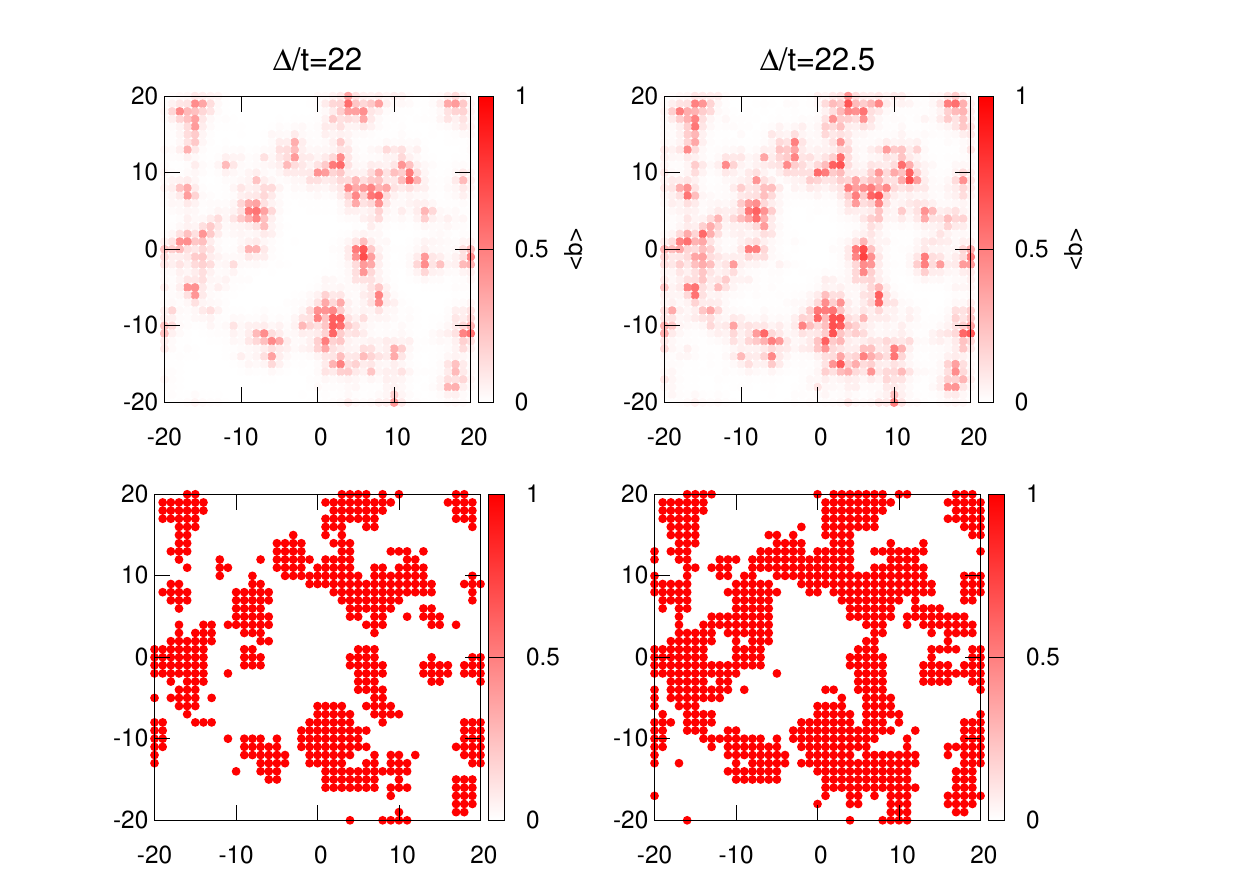}
\caption{(Colour online) Before and after spanning of a superfluid cluster as the disorder strength ($\Delta$) is increased keeping the disorder configuration fixed. The upper panel shows the actual $\langle{b\rangle}$ distribution in the the lattice both before (left) and after (right) the appearence of a spanning cluster. The lower panel shows the cluster distribution after standardising with a cut-off value for $\langle{b\rangle}$ ($>0.05$). The left panel corresponds to $\Delta/t=22$ and the right panel corresponds to $\Delta/t=22.5$ at $U/t=47$.}
\label{span}
\end{figure}

In the presence of uncorrelated box-like disorder, superfluid and Mott phase does not share a common boundary (Ref. \cite{3DdbhmQMC}). The Mott and superfluid phases are always mediated by the bose-glass phase. As discussed earlier, it is rather easier to distinguish between Mott phase and Bose-glass within the scope of mean-field self consistency. The self-consistently determined superfluid order parameter or $\langle{b\rangle}$ averaged over the system and disorder configurations clearly distinguishes Mott and Bose-glass phase. In Mott phase $\langle{b\rangle}$ is exactly zero and in Bose-glass phase $\langle{b\rangle}$ is small but non-zero. 

\section{distinguishing superfluid from BG : Percolation driven transition}

Mean-field self-consistant and disordered averaged $\langle{b\rangle}$ is not enough to distinguish the superfluid phase from the Bose-glass phase as in both the phases superfluid order parameter $\langle{b\rangle}$ is non-zero. Response functions which are non-local in space or time are impossible to compute within this mean-field scheme described above. Hence, the Edward Anderson's order parameter \cite{EdwardAndersons} for determination of glassy phase or the superfluid stiffness for determination of the superfluid phase are not suitable for detecting the  superfluid to Bose-glass transition within this mean-field self-consistency as these are limiting cases of response functions non-local in time. We understand the superfluid phase emerging out of the Bose-glass phase as a percolation driven transition, as widely discussed in the literatures previously \cite{KSheshadri,Niederle_2013,Nabi2016,Perc2}.

We incorporate a straight forward analysis which deals with the spanning of the clusters with non-zero order parameter ($\langle{b\rangle}$) across the sample. Within mean-field, the self-consistent order parameter ($\langle{b\rangle}$) between different superfluid clusters in a configuration typically do not have any phase fluctuations \cite{Nandini1993}. Hence, in further analysis we only take into account the absolute value of the order parameter. In order to disinguish between superfluid and Bose-glass phases we used Hoshen-Kopelman algorithm \cite{HKalg} to identify the clusters with non-zero order parameter ($\langle{b\rangle}$ greater than a cutoff) and determine if there exists a spanning cluster for a given disorder configuration. In Fig. \ref{span} (lower panel) we show $\langle{b\rangle}\ne0$ clusters with ($\Delta/t=22.5$) and without ($\Delta/t=22$) presence of a spanning cluster for a given disorder configuration at $U/t=47$. Fig. \ref{span} upper panel shows the actual $\langle{b\rangle}$ distribution before and after spanning occurs. If at least one superfluid cluster exists for a given set of parameters, spanning the length of a thermodynamically large system, that should equivalently imply non-zero superfluid stiffness. If self-averaging is performed in the system by averaging over a large number of disorder realizations it equivalently shows the physics of thermodynamically large system. For a given system size, we define a number $\mathcal{C}$ which is zero if there is no spanning cluster of non-zero $\langle{b\rangle}$ and one if there is a spanning cluster. Averaging over a large number of disorder realisations ($\langle{\mathcal{C}\rangle}$) for the same system size and parameters, we calculate the probability of occurence ($P_{span}=\langle{\mathcal{C}\rangle}$) of $\mathcal{C}=1$. The same procedure is carried out for different system sizes, then the distribution $P_{span}$ is plotted for different system sizes, over a region of the tuning parameter ($U$ or $\Delta$). The distribution $P_{span}$ for different system sizes  cross at a specific value of the tuning parameter which is the critical value of that parameter for the superfluid to Bose-glass transition. By using this method we are exploring the classical percolation transition of the sites with non-zero order parameter $\langle{b\rangle}$ determined by mean-field self-consistency. Hence presence of a spanning cluster, averaged over a large number of disorder realisation distinguishes the superfluid from the bose-glass phase. We explore the crossing of $P_{span}$ for different system sizes with the parameters of the phase diagram, $U$ or $\Delta$. 

In our calculations, the cut off value of $\langle{b\rangle}$ for a given site to determine if that site is a part of the a superfluid cluster or not is taken to be 0.05 for all $\Delta$ and $U$ values and system sizes. Varying the cut-off here does not alter the phenomenological picture and hence the general features of the phase diagram remains the same. The superfluid region in the final phase diagram will be larger or smaller in size for slight alteration of the cut-off, but the finger like re-entrant superfluid phases as predicted by the QMC calculations remain robust (discussed in detail in the following section). 

\begin{figure}[!b]
\includegraphics[width=8cm]{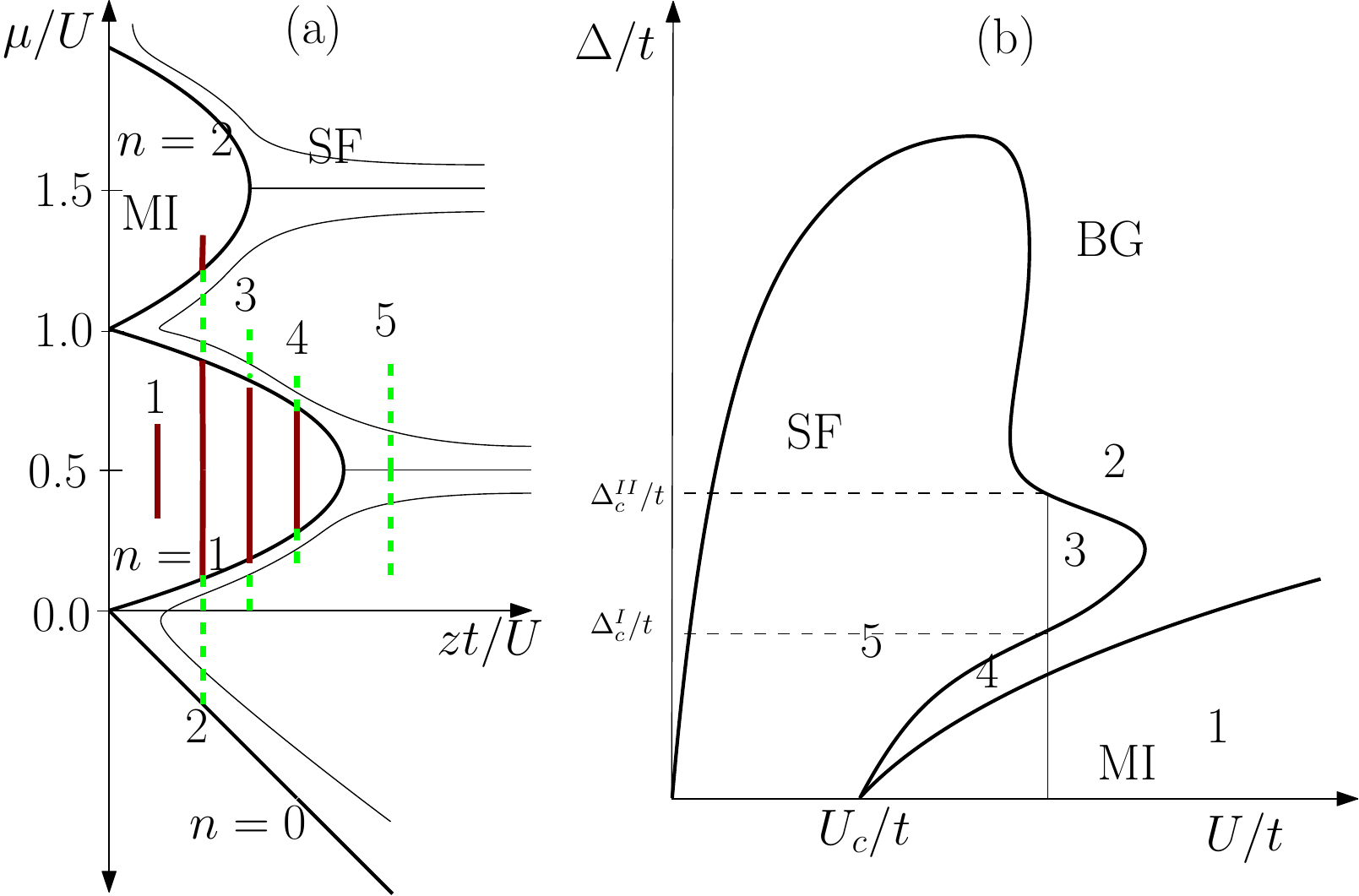} \\
\includegraphics[width=8cm]{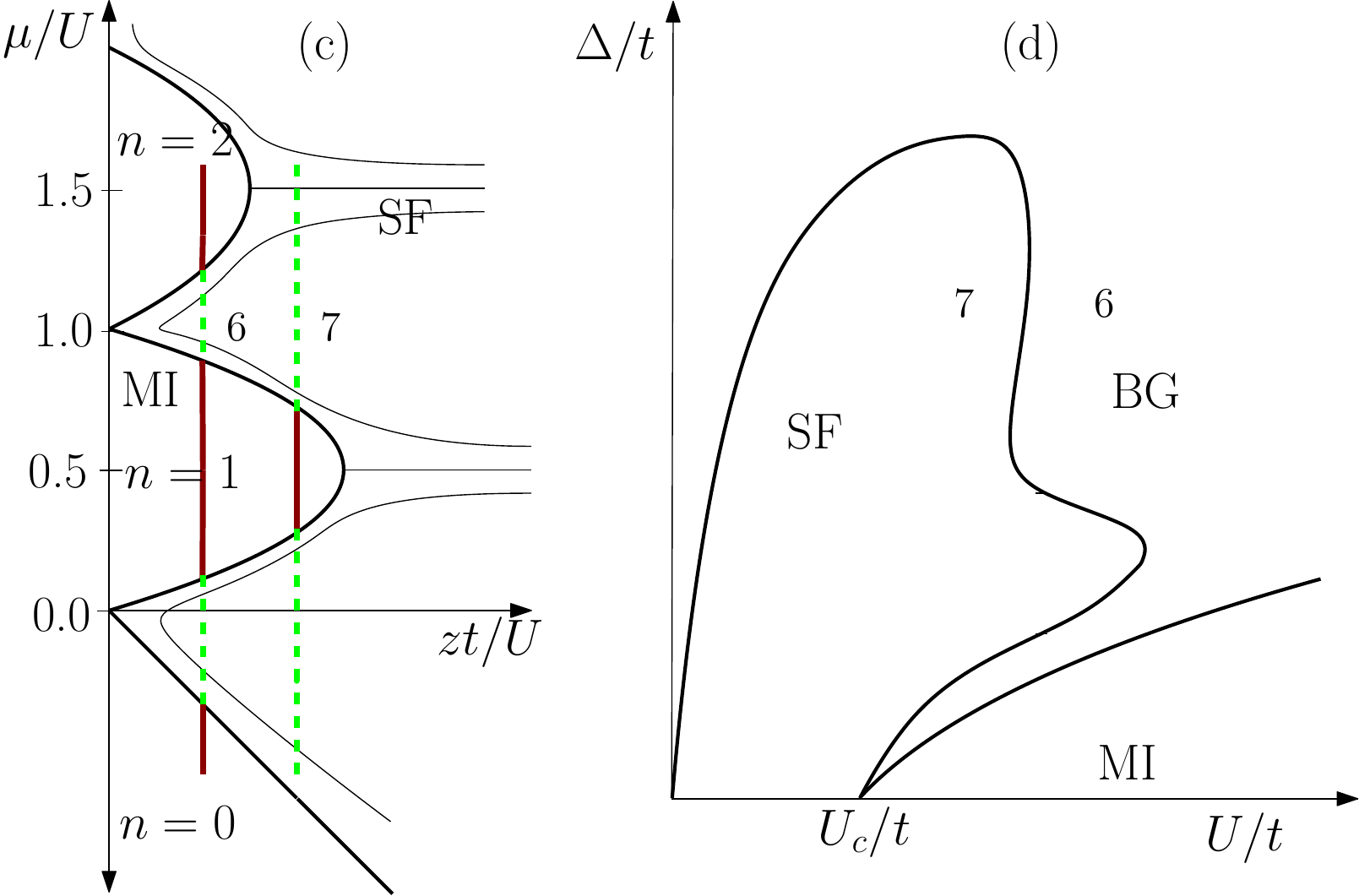}
\caption{(Colour online) Schematic phase diagram of $(a)$,$(c)$ Bose-Hubbard model and $(b)$,$(d)$ Disordered BHM for fixed filling $(n=1)$. The x-axis for $(a)$,$(c)$ ($zt/U$) and $(b)$,$(d)$ ($U/t$) is inverse of one another multiplied by a constant ($z$). Phenomenological understanding of $(b)$,$(d)$ $\Delta/t-U/t$ phase diagram for the disordered BHM can be obtained from the $(a)$,$(c)$ $\mu/U-t/U$ phase diagram for the pure case. Disorder strengths ($\Delta/U$) given by coloured line segments ($1$,$2$,$3$,$4$ and $5$) in $(a)$ corresponds to the respective cases in $(b)$ and coloured line segments $6$,$7$ corresponds to the respective cases in $(d)$. Red solid line denotes the part of the line segment in Mott phase and green dashed line denotes the part of the line segment in superfluid phase (within single-site mean-field approximation). $\Delta_c^{I}/t$ and $\Delta_c^{II}/t$ gives the critical disorder strength for re-entrant superfluid phase at a given $U/t$. In $(d)$ we show percolation transition as $\Delta/t$ is kept fixed and $U$ varied. $6$ and $7$ are two points indicating before and after of the transition. In $(c)$ the corresponding cases are shown by the line segments of length $\Delta/U$ and its position with respect to the Mott loabs. The detailed arguments for the correspondence are provided in the text.}
\label{Sch_phsd}
\end{figure}

\subsection{Predicted phase diagram from phenomenology at high $U$ and low $\Delta$}

Before discussing the mean-field results, it is important to revisit the pure Bose-Hubbard model and the phenomenology as disorder is introduced at fixed filling. As discussed above, previous simulational works have shown that the $\Delta/t-U/t$ phase diagram shows re-entrance of superfluid phase as $\Delta$ is increased for a fixed $U$ ($U>U_c$). The re-entrant superfluid phenomenon (at high $U$ and low $\Delta$) can be explained from the BHM phase diagram as discussed below. The pure BHM phase diagram in the $t/U-\mu/U$ plane has Mott lobes with different integral fillings and a vacuum region for negative chemical potential (Fig. \ref{Sch_phsd}$(a)$). The x-axis of Fig. \ref{Sch_phsd}$(a)$ ($zt/U$) refers to the inverse of the x-axis of Fig. \ref{Sch_phsd}$(b)$ ($U/t$) times the co-ordination number $z$. The filling is singular in the entire $\Delta/t-U/t$ phase diagram we are concerned with (Fig. \ref{Sch_phsd} $(b)$). The $\langle{n\rangle}=1$ contour which is a straight line parallel to the $t/U$ axis in the superfluid side meets the at the tip of the $\langle{n\rangle}=1$ Mott lobe \cite{Fisher}.

For pure BHM at fixed $n=1$ filling as one increases the on-site repulsion $U/t$ one approaches the first Mott lobe from right to left through the tip of the Mott lobe along the $\langle{n\rangle}=1$ contour (Fig. \ref{Sch_phsd} $(a)$). A line-segments of length $\Delta/U$ along the $\mu/U$ axis at fixed $zt/U$ gives the spread in the effective chemical potential for infinite system size. For simplicity we assume that the Mott lobe is symmetrical about the tip. Hence, centering the line-segments of length $\Delta/U$ at $\mu/U=0.5$ keeps the total filling to be approximately $\langle{n\rangle}=1$ even for $t/U$ in the superfluid region. However the actual phase diagram for BHM is not symmetric about the tip and the chemical potential at the center of the line segment (of length $\Delta/U$), which keeps the filling singular, may vary from $\mu/U=0.5$ for different disorder configurations and for a finite system size. For pure BHM, as we approach the $\langle{n\rangle}=1$ contour from right to left (Fig. \ref{Sch_phsd} $(a)$), at a critical $U_c$ (or $t/U_c$) Mott region ($n=1$) starts. 

For $U>U_c$ (or $t/U < t/U_c$), if the $\Delta$ is small ($\Delta/U << 1$) such that the spread of chemical potential centered around $\mu/U=0.5$ resides entirely within the $n=1$ Mott lobe (case-$1$ in Fig. \ref{Sch_phsd} $(a)$) then we have Mott region in the $\Delta/t-U/t$ phase diagram (case-$1$ in Fig. \ref{Sch_phsd} $(b)$). As $\Delta$ is increased further (for $U>U_c$), parts of the chemical potential spread would reside in the superfluid region. For a given disorder configuration, the fraction of sites in the superfluid region depends on the ratio between the effective chemical potential spread  inside and outside the Mott lobe. For our phenomenlogical analysis, we assume that the value of $\langle{b\rangle}$ at a given site is independent of that of the neighbouring sites (i.e. single site self-consistency; in Eq \ref{eq6}, $\phi_i=zt\langle{b\rangle}_i$). Then by the ratio of the length of the line segment in green dashed  (Fig. \ref{Sch_phsd} $(a)$) to the total length of the line segment, i.e. red solid + green dashed in Fig \ref{Sch_phsd}$(a)$ one can represent the tuning parameter for classical percolation. If the fraction of sites for non zero $\langle{b\rangle}$, greater than a critical value (percolation threshold), a spanning cluster can occur. Hence, there can be phase coherence across the sample and a superfluid phase for $\Delta > \Delta_c^I$ (case $3$).

\begin{figure*}[!ht]
\includegraphics[width=11cm]{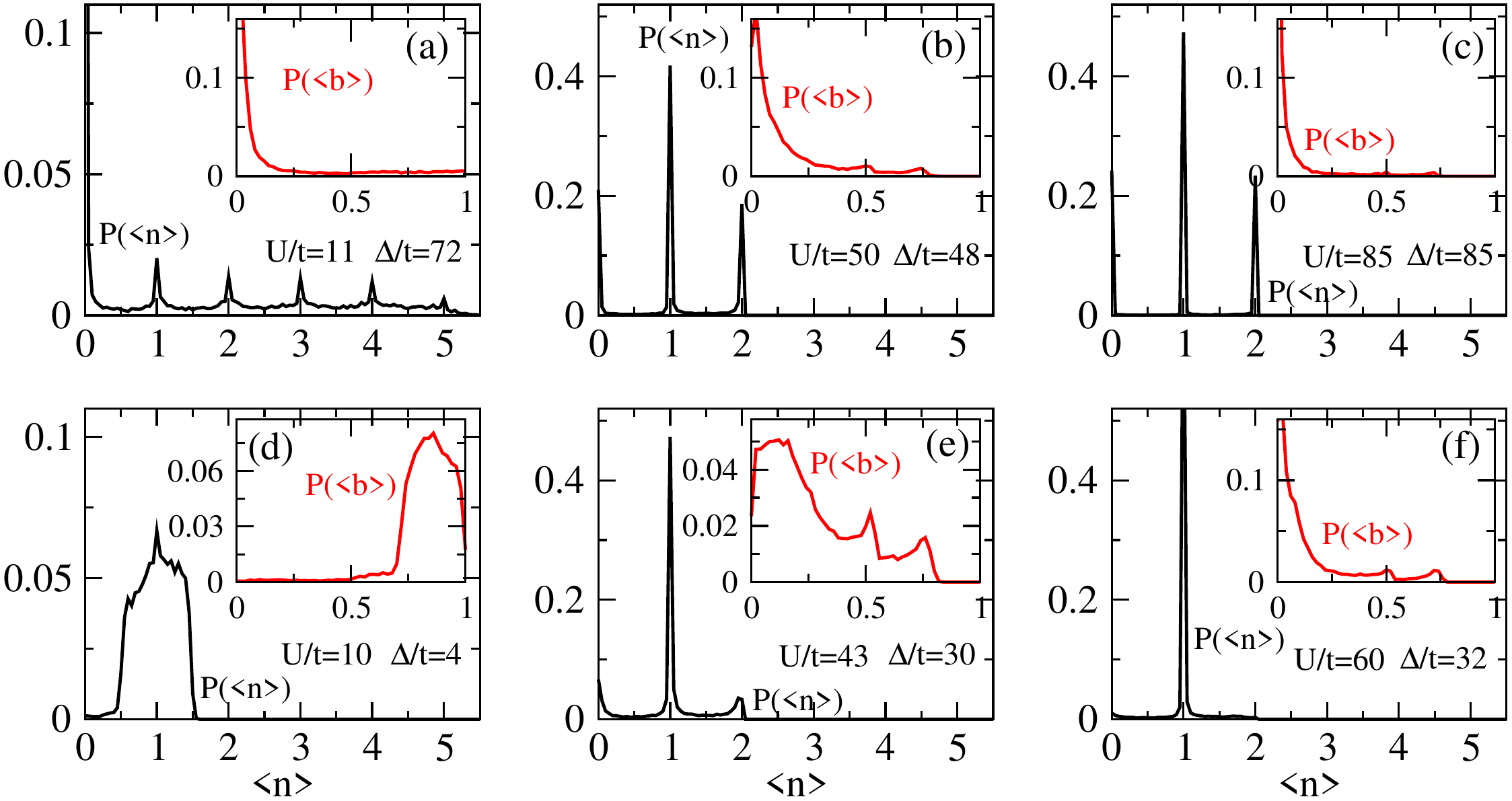}
	\caption{(Colour online) Probability distribution of the order parameter $\langle{b\rangle}$ as well as the probability distribution for the density $\langle{n\rangle}$ (inset) at different parts of the phase diagram ($(a)$ to $(f)$). The different parameters ($(a)$ to $(f)$) corresponds to different points in the phase diagram as shown in Fig. \ref{phsd_dbhm}.}
\label{BavNav}
\end{figure*}

\begin{figure}[!hb]
\includegraphics[width=10cm]{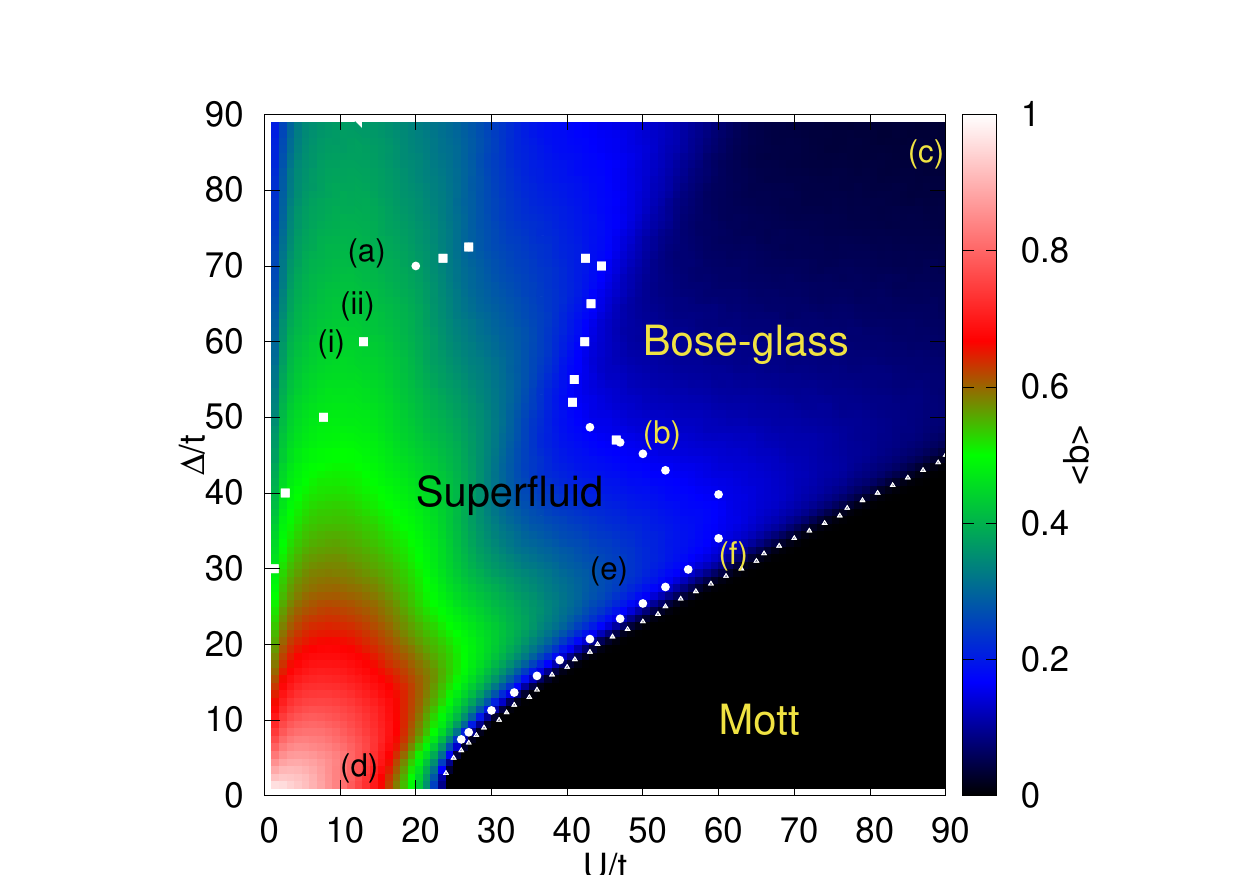}
	\caption{(Colour online) $\Delta/t-U/t$ phase diagram for DBHM at fixed filling of $\langle{n\rangle}=1$. The disorder averaged mean-field order parameter $\langle{b\rangle}$ is plotted with a false colour plot. We have used $l=15$ and averaged over 300 disorder configurations. Solid white squares and circles give the boundary of the superfluid and Bose-glass phase. Solid white triangles give the boundary for Mott and Bose-glass phase. $(a)$-$(f)$ indicate the parameter points for Fig. \ref{BavNav}. Also, $(i)$ and $(ii)$ indicate the parameter points for Fig. \ref{lowu}}
\label{phsd_dbhm}
\end{figure}

As we further increase $\Delta$ (for $U>U_c$) the spread of effective chemical potential also includes the $\langle{n\rangle}=2$ Mott lobe and the vacuum ($\langle{n\rangle}=0$), in that case the aforementioned ratio decreases and percolating clusters spanning the whole system cease to exist at a critical disorder strength ($\Delta>\Delta_c^{II}>\Delta_c^I$). This is case-$2$ as in Fig \ref{Sch_phsd}$(a)$. As we increase $U$ further, there exists a maximum value of $U$ (at the tip of the fingurelike region) such that for values of $U$ greater than that the re-entrant superfluid phase is absent for all values of $\Delta$. This expalins the finger-like re-entrance behavior of superfluid (case-$4$,$3$ and $2$). In Fig \ref{Sch_phsd}$(a)$ the red (solid) region of the line segments are inside the Mott region and the green (dashed) regions are inside the superfluid region. Different line segments in Fig. \ref{Sch_phsd}$(a)$  represent different points in the $\Delta-U$ phase diagram in Fig. \ref{Sch_phsd}$(b)$ as given by the corresponding numbers. An important point to note - in the above discussion, changing $\Delta$ keeping $U$ fixed equivalently imply changing the line segment of length $\Delta/U$ in Fig. \ref{Sch_phsd}$(a)$ at a fixed $zt/U$.

In Fig \ref{Sch_phsd}$(c)$ and Fig \ref{Sch_phsd}$(d)$ percolation driven transition at a fixed value of $\Delta$ while $U$ is being varied is explained. As $U$ is varied, both the y-axis in Fig \ref{Sch_phsd}$(c)$ as well as the line segment corresponding to cases $6$ and $7$ parallel to the y-axis changes as $1/U$. Hence, increasing $U$ keeping $\Delta$ fixed effectively implies moving towards negetive x-direction. The ratio of green (dashed) to green (dashed)+red (solid) region is smaller for case-$6$ than for case-$7$. This implies that a percolation driven transition is possible between points $6$ and $7$ in Fig \ref{Sch_phsd}$(d)$.

The discussion above clearly reveals that within the phenomenological understanding of single-site strong-coupling mean-field theory and percolation transition the re-entrant superfluid regions for singular filling DBHM can be easily explained.

\begin{figure*}[!ht]
\includegraphics[width=14cm]{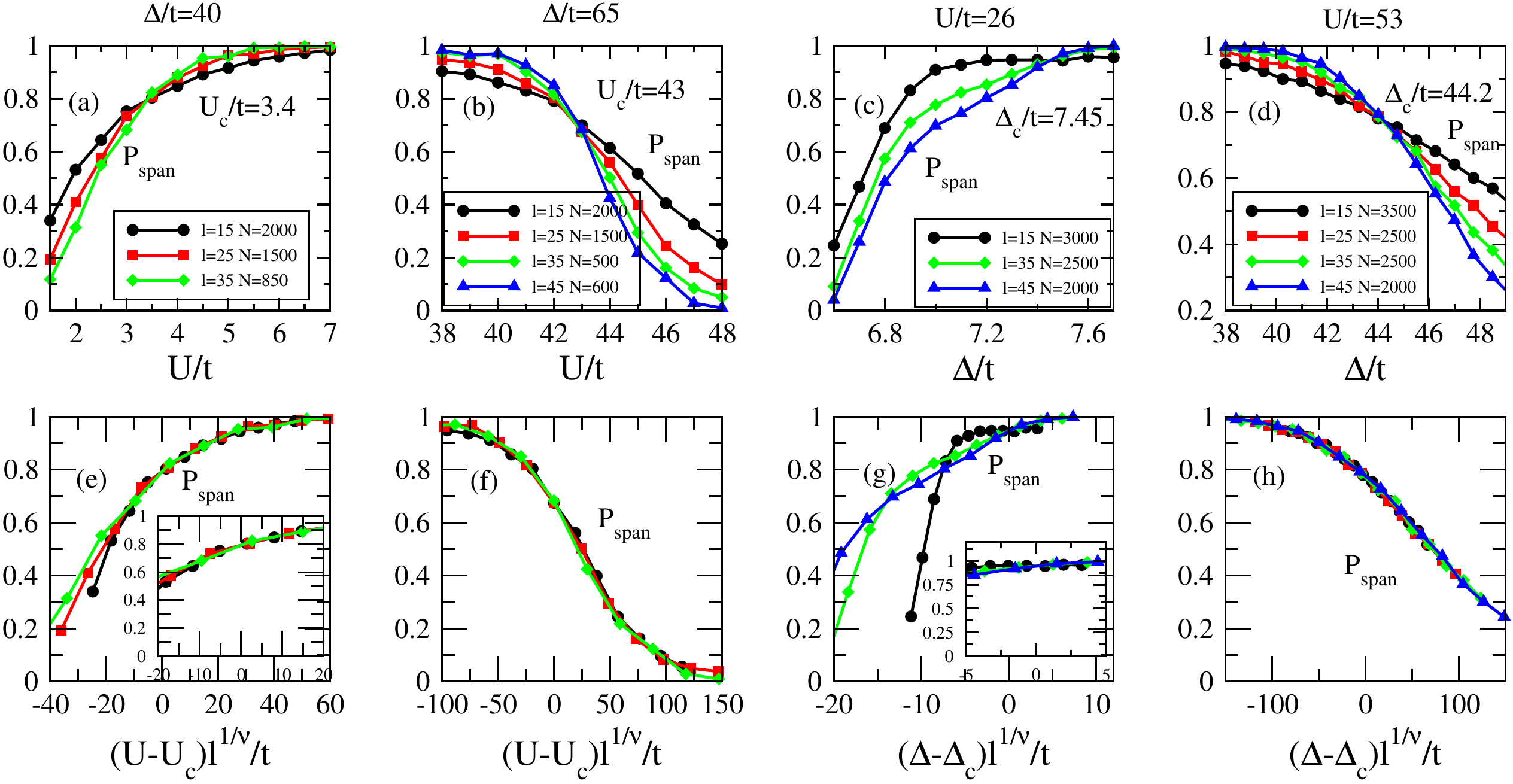}
	\caption{(Colour online) Upper panel: $P_{span}$ is plotted for different system sizes for fixed $U/t$ $(a)$ $\Delta/t=40$, $(b)$ $\Delta/t=65$, and fixed $\Delta/t$ $(c)$ $U/t=26$, $(d)$ $U/t=53$. Lower panel: data collapse for $(e)$ $\Delta/t=40$, $(f)$ $\Delta/t=65$, $(g)$ $U/t=26$, $(h)$ $U/t=53$. Inset plots in $(e)$ and $(g)$ show a zoomed in view of data collaspe near the transition point.}
\label{Datacoll}
\end{figure*}

\subsection{Phase diagram at fixed filling $\langle{n\rangle}=1$: Results}

For the ground state phase diagram, we self-consistently determine the order parameter $\langle{b\rangle}$ for a $15 \times 15$ lattice, with periodic boundary condition and $300$ disorder realizations for different values of interaction strength and disorder strength. Both the interaction strength ($U/t$)  and the disorder strength ($\Delta/t$) are varied from $1$ to $90$. As parameters with $U/t<1$ is beyond the scope of strong-coupling mean-field, the $U \rightarrow 0$ limit has not been explored. The self-consistently determined order parameter $\langle{b\rangle}$ for a given value of on site repulsion $U/t$ and disorder strength $\Delta/t$ is averaged over all sites (for a given disorder realisation) as well as number of disorder configurations. In Fig. \ref{phsd_dbhm} we show $\langle{b\rangle}$ in a false colour plot, black region indicating $\langle{b\rangle}=0$ (Mott phase). Disorder averaged $\langle{b\rangle}$ can only distinguish between Mott and the superfluid (or Bose-glass) phase, as for both BG and SF the disorder averaged order parameter is non-zero. In the absence of disorder the critical interaction strength for superfluid to Mott transition occurs at $U/t=24$ within mean field \cite{Fisher,KSheshadri}. As we increase the disorder for a fixed $U/t$ ($>24$) the Mott phase survives upto a critical disorder strength. (A phenomenological understanding is presented in the previous section.) With increasing values of $U/t$ the critical $\Delta/t$ also monotonically increases. In Fig \ref{phsd_dbhm}, the Mott to superfluid boundary is shown by the white triangles, indicating the the point where the disorder averaged $\langle{b\rangle}$ goes to zero exactly.

Fig. \ref{BavNav} shows distribution of $\langle{b\rangle}$ and $\langle{n\rangle}$ at various regions of the phase diagram given in Fig. \ref{phsd_dbhm}. Deep into the BG phase Fig. \ref{BavNav} $(c)$, probability of $\langle{b\rangle}$ is peaked at zero with a very small spread in the non-zero values and the $\langle{n\rangle}$ is peaked at the integral values. Fig. \ref{BavNav} $(b)$ shows a parameter point just outside the high $U$ finger-like region. Here there is a significant spread of $P(\langle{b\rangle})$ for non-zero $\langle{b\rangle}$ and $\langle{n\rangle}$ is peaked at $n=1$ as well as $n=2$ and $n=0$ indicating presence of double occupancy and null occupancy Mott clusters. Fig. \ref{BavNav} $(d)$ shows a case deep in the superfluid region, for low $\Delta$ values, where cluster of sites in Mott phase are clearly absent as $P(\langle{b\rangle})$ is zero for $\langle{b\rangle}=0$. Both $P(\langle{b\rangle})$ and $P(\langle{n\rangle})$ indicate that the whole lattice is in the supefluid phase for any disorder configuration. Fig. \ref{BavNav} $(a)$, at low values of $U$ and high values of $\Delta$ one can observe $P(\langle{n\rangle})$ peaks at integral fillings as well as $P(\langle{b\rangle})$ is nonzero for nonzero values of $\langle{b\rangle}$ along with a peak in $\langle{b\rangle}=0$. In Fig. \ref{BavNav} $(f)$, $P(\langle{n\rangle})$ is peaked only at $n=1$, but $P(\langle{b\rangle})$ has a spread in the nonzero $\langle{b\rangle}$. Fig. \ref{BavNav} $(e)$ shows a region indicated in Fig. \ref{phsd_dbhm}, inside the `fragile' re-entrant superfluid region. Here the distribution $P(\langle{b\rangle})$ has significant contribution for $\langle{b\rangle}\ne0$ which is different from the case Fig. \ref{BavNav} $(d)$. Both the distributions $P(\langle{b\rangle})$ and $P(\langle{n\rangle})$ change continuously throught the phase diagram as the parameters are varied.

The phase boundary of superfluid region and Bose-glass region is obtained from the percolation analysis as discussed previously. We have studied the percolation driven bose-glass to superfluid transition by exploring the statistics of spanning clusters of non-zero superfluid order parameter. In Fig. \ref{phsd_dbhm} we distinguish the Bose-glass and the superfluid phase by the crossing point of $P_{span}$ for different system sizes as $\Delta$ or $U$ is being tuned. In Fig \ref{Datacoll} we show crossing of $P_{span}$ for different system sizes and for various parameter ranges throughout the phase diagram. Some previous works have also used percolation analysis for determination of the BG to SF transition \cite{Niederle_2013,Nabi2016}. Based on the mean-field result, we extensively calculated the percolation probability at various values of $U$ and $\Delta$, at different parameter region, and varied either $U$ or $\Delta$ (keeping the other fixed) to perform the finite size scaling analysis to determine the phase diagram. In Fig \ref{phsd_dbhm}, the boundary points given by solid white circles are obtained by exploring the superfluid to Bose-glass transition by varying $\Delta$ at a fixed $U$ and solid white squares indicate the transition between SF and BG where the transition point is obtained by varying $U$ at a fixed value of $\Delta$. Fig. \ref{Datacoll}, upper panel shows the crossing of $P_{span}$ curve for different system sizes where $\Delta$ or $U$ being varied. Fig. \ref{Datacoll} lower panel, $P_{span}$ is plotted for scaled parameters ($\Delta$ or $U$). Fig \ref{Datacoll} $(e)$,$(f)$ for $\Delta/t=40$ and $\Delta/t=65$ as a function of $(U-U_c)l^{1/\nu}/t$ and Fig \ref{Datacoll} $(g)$,$(h)$ for $U/t=26$ and $U/t=53$ as a function of $(\Delta-\Delta_c)l^{1/\nu}/t$ respectively. The inset plots in Fig \ref{Datacoll} $(e)$ and $(g)$ show the scaling region more clearly where the data collapse is more identifiable, indicating smaller scaling region as compared to other parts of the phase diagram (Fig \ref{Datacoll} $(f)$ and $(h)$).

The phase diagram as shown in the Fig. \ref{phsd_dbhm}  has two re-entrant finger-like superfluid regions also referred as `fragile' SF region (\cite{3DdbhmQMC,2DdbhmQMC}). One at the higher values of $U$ and lower values of $\Delta$ and the other at very high values of $\Delta$ and low $U$. In these finger-like regions, superfluidity arises because of interplay between the interaction and disorder. The region outlined with white dots and white squares represent the superfluid region with at least one spanning cluster of non-zero $\langle{b\rangle}$ for infinite system size within mean-field analysis. For large $U$ ($U/t>24$) as $\Delta$ is increased, beyond a critical disorder ($\Delta_c^I$), spanning clusters can occur hence phase coherence across the sample is possible. Increasing $\Delta$ further leads to more sites having $\langle{n\rangle}=2$ Mott phase and vacuum ($\langle{n\rangle}=0$). Probability of having SF regions goes below the percolation threshold. Beyond a critical $U$, as we increase $\Delta$, the percolation threshold is never achieved. For low $U$ and high $\Delta$ superfluid phase appears as percolation driven transition as $U$ is increased keeping $\Delta$ fixed. As $U$ is increased further keeping $\Delta$ fixed, a percolation driven superfluid-BG transition occurs. According to previous studies \cite{3DdbhmQMC}, the Mott to superfluid transition is replaced by Mott to Bose-glass to superfluid, which we also observe in our result (Fig. \ref{phsd_dbhm}) as there exists a small region of Bose-glass phase in-between the Mott and the superfluid phase. The two dimensional phase diagram obtained from mean-field and percolation analysis compares qualitatively well with the QMC results as obtained in Ref. [\onlinecite{2DdbhmQMC}]. The tip of the fingurelike superfluid region at low $\Delta$ and high $U$ appears approximately around $U/t \approx 62$. The QMC result for the aforementioned tip of the fingurelike region is at $U/t=49$. The previous works based on percolation analysis \cite{Nabi2016}, also produced similar phase diagram. However, we do not find exact quantitative agreement between the phase diagram as given in Ref. [\onlinecite{Nabi2016}] and our result. 

We study the data collapse for the $P_{span}$ as a function of scaled argument using appropriate scaling exponents for classical percolation in two dimensions as also used in Ref. \cite{Niederle_2013}. With the classical percolation exponents for two dimensions ($\nu=1.33$) \cite{2DpercBook}, data-collapse can be observed over a region of parameter space near the BG-SF transition point. We observed the scaling region to be shorter (in the parameter range) for the Bose-glass to superfluid transition where the transition point is close to a Mott-SF transition at low values of $\Delta$. In Fig. \ref{Datacoll} data collapse is demonstrated in the lower panel. It shows data collapse for the transition along both the $U$ direction and $\Delta$ direction respectively. Along the x-axis the appropriate argument for the scaling function is plotted with appropriate values of $U_c$ or $\Delta_c$ obtained from the $P_{span}$ calculations. \cite{Ceperley}.

\begin{figure}[!b]
\includegraphics[width=9cm]{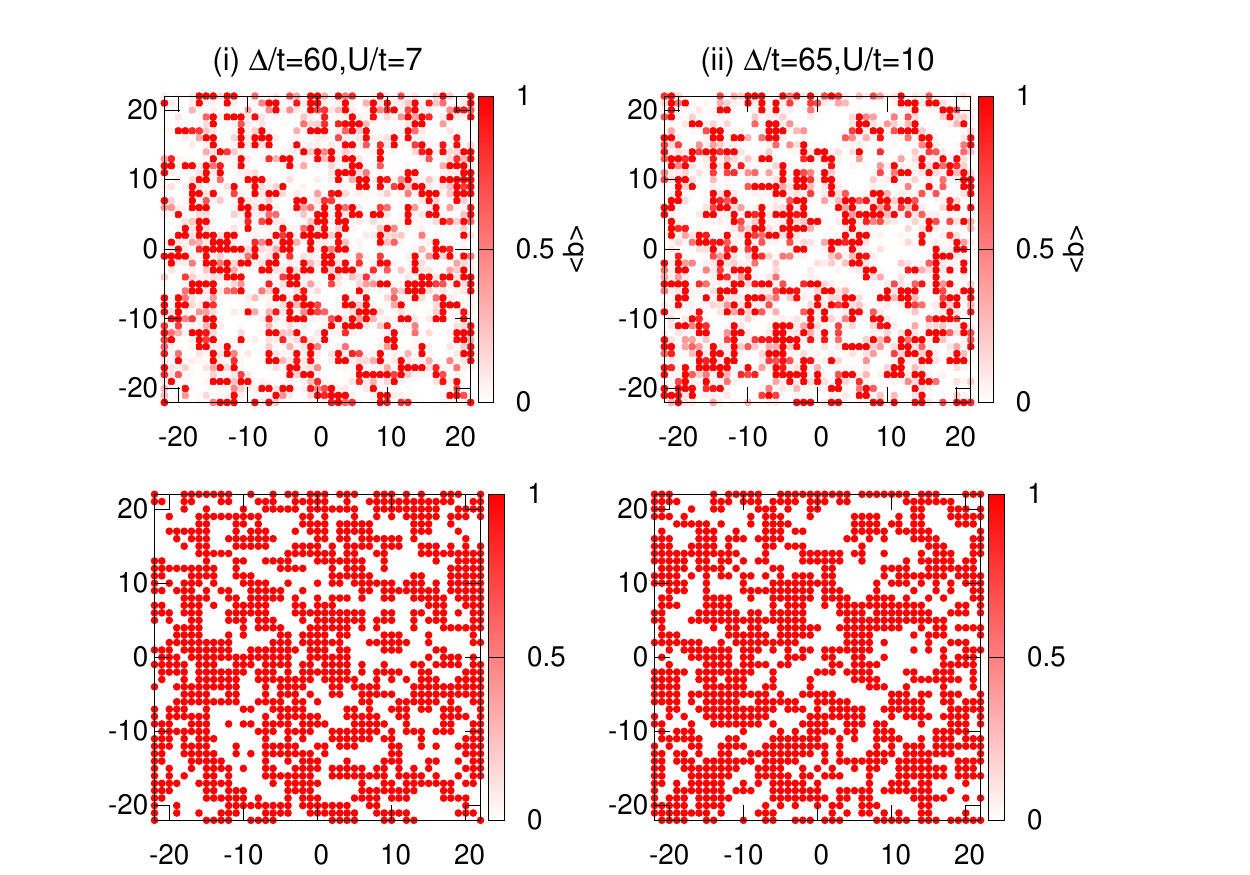}
	\caption{(Colour online) The lower panel showing absence of spanning cluster at low $U$ and high $\Delta$ where $\langle{b\rangle}$ values are high. The left panel is for $(i)$$\Delta/t=60$, $U/t=7$ and the right panel is for $(ii)$$\Delta/t=65$, $U/t=10$. In Fig \ref{phsd_dbhm} the above mentioned parameter regions are shown. }
\label{lowu}
\end{figure}

\subsection{Some remarks on the spanning clusters}

Within the mean-field picture the spanning superfluid clusters give some understanding of the phase diagram of disordered BHM. The probability ($P_{span}$) of a spanning cluster shows crossing for different system sizes indicating the superfluid-bose glass transition. However, the cluster of sites with non-zero superfluid order parameter does not appear randomly within the sample. The formation of clusters of non-zero $\langle{b\rangle}$ is correlated as indicated in Eq. \ref{eq6}. The self-consistent $\langle{b\rangle}$ at a given site is a sum of $\langle{b\rangle}$ in the nearest neighbour sites. Hence, it can be termed as formation of liquid drops \cite{Droplet2008}. 

In Fig. \ref{lowu} we show two situations where the average order parameter $\langle{b\rangle}$ throughout the sample is significantly high although no spanning cluster of non-zero $\langle{b\rangle}$ exists. The left panel (Fig. \ref{lowu} $(i)$) shows $\Delta/t=60$, $U/t=7$ and the right panel (Fig. \ref{lowu} $(ii)$) shows $\Delta/t=65$, $U/t=10$. For both the cases the upper panel shows the actual order parameter profile throughout the lattice and the lower panel shows the cluster pattern derived from the upper panel using a cut off. These two cases are obtained from low $U$ and high $\Delta$ region of Fig. \ref{phsd_dbhm}, where the false colour plot of disorder averaged $\langle{b\rangle}$ show higher values on the BG region than the superfluid region. This clearly shows, simple contour plots of disorder averaged $\langle{b\rangle}$ does not match with phase boundary obtained from percolation transition analysis. 
 
In our percolation analysis, we have not considered the corner sharing clusters to be the same cluster as the self-consistency condition (Eq. \ref{eq6}) only takes into account the nearest neighbouring sites. In Ref. \cite{Droplet2008} it is argued, that the superfluid clusters have a droplet like feature due to spatial correlation \cite{CorrDis}. A more rigorous percolation analysis taking into account this general tendency of the superfluid clusters to form droplets can provide a better understanding of the transition within this picture. Also, cluster analysis for the different Mott phases ($n=1$ and $n=2$, and vaccum $n=0$) deep inside the Bose-glass phase may provide useful understanding of the different forms of excitations possible there.

\section{Conclusion}

In conclusion we perform the mean-field and classical percolation analysis to distinguish various phases of disordered BHM at fixed $\langle{n\rangle}=1$ filling. Each point at the boundary of the SF-BG transition is obtained from the crossing point of the spanning cluster probability ($P_{span}$) for different system sizes, across the SF-BG crossing, varying either $\Delta$ or $U$. We also argue the importance of mean-field along with the percolation analysis methods in developing understading of re-entrant superfluid region and bose-glass phase at high $U$ low $\Delta$.

\section{Acknowledgements}

MG acknowledges useful discussions with Dr. Arijit Dutta Dr. Yogeshwar Prasad and Prof. Pinaki Majumdar. MG also acknowledges HPC cluster facilities of Harish-Chandra Research Institute, Allahabad.

\bibliography{dBHM_writeup.bib}
\end{document}